\documentclass[preprint2]{aastex}

\shortauthors{Wright \etal}

\shorttitle{WISE Mission}

\newcommand{\etal}         {{\it et al.}}

\newcommand{\be}           {\begin{equation}}
\newcommand{\ee}           {\end{equation}}
\newcommand{\bea}          {\begin{eqnarray}}
\newcommand{\eea}          {\end{eqnarray}}

\begin{document}

\title{A WISE Observation of a coolest brown dwarf, CFBDSIR 1458+1013.}


\author{
Edward L.\ Wright\altaffilmark{1},
Amy Mainzer\altaffilmark{2},
Chris Gelino\altaffilmark{3},
Davy Kirkpatrick\altaffilmark{3}
}

\altaffiltext{1}{UCLA Astronomy, PO Box 951547, Los Angeles CA 90095-1547}
\altaffiltext{2}{Jet Propulsion Laboratory, 4800 Oak Grove Drive, Pasadena, CA 91109}
\altaffiltext{3}{Infrared Processing and Analysis Center,
California Institute of Technology, Pasadena CA 91125}

\email{wright@astro.ucla.edu}

\begin{abstract}
The Wide-field Infrared Survey Explorer (WISE) has 
detected the close binary brown dwarf system
CFBDSIR 1458+1013AB as WISEP J145829.35+101341.8
with a combined magnitude at 4.6 $\mu$m of W2 = $15.488 \pm 0.147$.
This allows a comparison with another ``coolest'' brown dwarf
candidate WD 0806-661B
that has been observed at 4.5 $\mu$m with [4.5] = $16.75 \pm 0.05$.
Here we use the WISE data to show that 1458+1013B is almost certainly warmer and
more luminous than WD 0806-661B.
\end{abstract}


\maketitle

\section{Introduction}

The Wide-field Infrared Survey Explorer (WISE) \citep{wright/etal:2010}
has surveyed the entire sky in four thermal infrared bands.
The Spitzer 4.5 $\mu$m band and the WISE 4.6 $\mu$m (W2) band are
very similar in wavelength, so no large color term is expected
when comparing magnitudes in these bands.
We have examined the set of spectroscopically confirmed T dwarfs
seen by both WISE and Spitzer and see only a small
color term, with mean [4.5]-W2 = 0.054 magnitudes and no
apparent trend with color or spectral type.

We can use this to estimate the W2 magnitude of WD 0806-661B
to be 16.7 based on the Spitzer data \citep{luhman/burgasser/bochanski:2011}, 
which is below the sensitivity limit for WISE.  
Given the 1.25 $\mu$m limit
of J $> 21.7$ \citep{rodriguez/etal:2011},  the color is
J-W2 $> 5.0$.  With the $19.2 \pm 0.6$ 
pc distance \citep{luhman/burgasser/bochanski:2011}, 
the absolute magnitude is $M_{W2} = 15.28$.

\section{Color-Magnitude Fit}

\begin{figure}[t]
\plotone{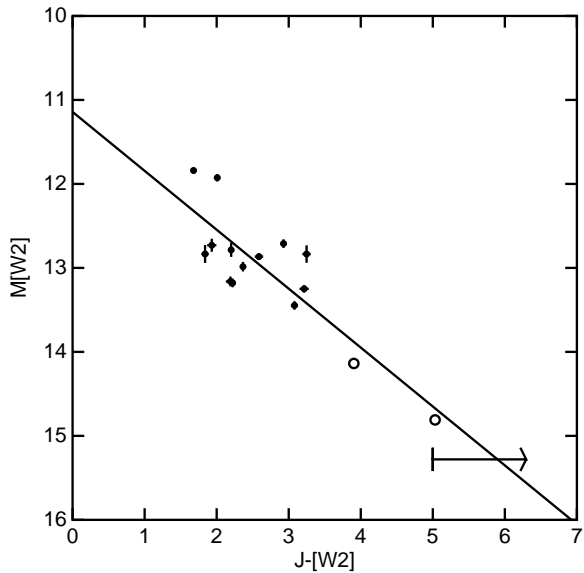}
\caption{A color magnitude diagram for T brown dwarfs.
The solid points with errorbars are T dwarfs \citep{patten/etal:2006}
while the open circles are 1458+1013 A \& B with the flux ratio
derived in the fit.  The rightward arrow in the lower right shows
WD 0806-661B \citep{luhman/burgasser/bochanski:2011,
rodriguez/etal:2011}\label{fig:CM}}
\end{figure}

The WISE data on the close binary 1458+1013 only give the
combined light at 4.6 $\mu$m, 
with a W2 magnitude of $15.488 \pm 0.147$.
The other WISE bands only give upper limits on the flux,
with a 2$\sigma$ limit on the combined magnitude at
3.4 $\mu$m of W1 $> 16.84$ magnitudes.
The brightness of the secondary component of the binary 
depends on the assumed flux
ratio $f = F_A/F_B$.  This ratio is 5.2, 8.6 \& 7.6 at 1.25,
1.6 \& 2.15 $\mu$m \citep{liu/etal:2011}, but we
expect the B component of the binary will be redder
than the A component leading to a lower flux ratio at
4.6 $\mu$m.  We have estimated the flux ratio at
4.6 $\mu$m by fitting a straight line
$M_{W2} = a+b(J-W2)$ to a sample consisting of
brown dwarfs with known distances \citep{patten/etal:2006}
plus the A and B components of 1458+1013 with the flux
ratio $f$ as a third parameter of the fit.  Figure \ref{fig:CM}
shows the best fit, which has $a = 11.14$ and $b = 0.701$.
The scatter is larger than can be explained
by observational errors, so the error on $f$ from the fit
is calculated assuming a intrinsic scatter of $\pm 0.42$ on
$M_{W2}$.  This gives a flux ratio $f = 1.85 \pm 0.61$.
For the best fit flux ratio 1458+1013B is as red as the lower limit
on the color for WD 0806-661B, but also considerably more luminous
at 4.6 $\mu$m.  For larger flux ratios the B component gets
fainter but also bluer.  If the straight line fit to $M_{W2}$ to $(J-W2)$ 
were exact then $f(W2) = f(J)^{b/(1+b)} = 1.97$.  For this
flux ratio the color of 1458+1013B is J-W2 = 5.01 and the absolute 
magnitude is $M_{W2} = 14.85$, insignificantly different from the 
plotted solution.

\section{Discussion}

Given that WD 0806-661B is  both 20\% closer to the Sun
and its flux is 3 times
fainter than the combined light of 1458+1013 at 4.6 $\mu$m,
our conclusion that  WD 0806-661B  is the cooler and less luminous
of these two ``coolest''
brown dwarfs is straightforward.  The same conclusion can be made in
the J band where the \citet{rodriguez/etal:2011} limit on 
WD 0806-661B is fainter than the measured
magnitude of 1458+1013B \citep{liu/etal:2011}.
This conclusion is also reflected in the estimated effective temperatures:
$370 \pm 40$~K for 1458+1013B \citep{liu/etal:2011}
and $\approx 300$~K for WD 0806-661B
\citep{luhman/burgasser/bochanski:2011}.
But the fitted line predicts that the color of WD 0806-661B is
J-W2 = $5.9 \pm 0.6$.  If this prediction is correct then the
apparent J magnitude of WD 0806-661B will be J = 22.6,
and followup spectroscopy
to confirm that this object is a brown dwarf
will be impossible using ground-based telescopes and
quite difficult even with the Hubble Space Telescope.
The WISE all-sky survey should find objects this red
and redder that are much closer to the Sun and thus much 
more suited for detailed study.

\acknowledgements
This publication makes use of data products from the Wide-field
Infrared Survey Explorer, which is a joint project of the University
of California, Los Angeles, and the Jet Propulsion Laboratory/California
Institute of Technology, funded by the National Aeronautics and
Space Administration.


\begin{thebibliography}{5}
\expandafter\ifx\csname natexlab\endcsname\relax\def\natexlab#1{#1}\fi

\bibitem[{{Liu} {et~al.}(2011){Liu}, {Delorme}, {Dupuy}, {Bowler}, {Albert},
  {Artigau}, {Reyle}, {Forveille}, \& {Delfosse}}]{liu/etal:2011}
{Liu}, M.~C., {Delorme}, P., {Dupuy}, T.~J., {Bowler}, B.~P., {Albert}, L.,
  {Artigau}, E., {Reyle}, C., {Forveille}, T., \& {Delfosse}, X. 2011, ArXiv
  e-prints

\bibitem[{{Luhman} {et~al.}(2011){Luhman}, {Burgasser}, \&
  {Bochanski}}]{luhman/burgasser/bochanski:2011}
{Luhman}, K.~L., {Burgasser}, A.~J., \& {Bochanski}, J.~J. 2011, ApJL, 730, 9

\bibitem[{{Patten} {et~al.}(2006){Patten}, {Stauffer}, {Burrows}, {Marengo},
  {Hora}, {Luhman}, {Sonnett}, {Henry}, {Raghavan}, {Megeath}, {Liebert}, \&
  {Fazio}}]{patten/etal:2006}
{Patten}, B.~M., et al.  2006, \apj, 651, 502

\bibitem[{{Rodriguez} {et~al.}(2011){Rodriguez}, {Zuckerman}, {Melis}, \&
  {Song}}]{rodriguez/etal:2011}
{Rodriguez}, D.~R., {Zuckerman}, B., {Melis}, C., \& {Song}, I. 2011, ArXiv
  e-prints

\bibitem[{{Wright} {et~al.}(2010){Wright}, {Eisenhardt}, {Mainzer}, {Ressler},
  {Cutri}, {Jarrett}, {Kirkpatrick}, {Padgett}, {McMillan}, {Skrutskie},
  {Stanford}, {Cohen}, {Walker}, {Mather}, {Leisawitz}, {Gautier}, {McLean},
  {Benford}, {Lonsdale}, {Blain}, {Mendez}, {Irace}, {Duval}, {Liu}, {Royer},
  {Heinrichsen}, {Howard}, {Shannon}, {Kendall}, {Walsh}, {Larsen}, {Cardon},
  {Schick}, {Schwalm}, {Abid}, {Fabinsky}, {Naes}, \&
  {Tsai}}]{wright/etal:2010}
{Wright}, E.~L., et al.  2010, \aj, 140, 1868

\end{thebibliography}


\end{document}